\documentstyle[aps,tighten,preprint]{revtex}
\newcommand{\lsim}{\mathrel{\mathop{\kern 0pt \rlap
  {\raise.2ex\hbox{$<$}}}
  \lower.9ex\hbox{\kern-.190em $\sim$}}}
 
\begin{document}

\draft
\preprint{
\begin{tabular}{r}
JHU--TIPAC 97003 \\
DFTT 15/97 \\
KIAS--P97002 \\
\end{tabular}
}

\title{Finite temperature effects on the neutrino decoupling
\\
in the early Universe}

\author{
N. Fornengo$^{\rm a,b}$
\footnote{E-mail: fornengo@jhup.pha.jhu.edu,$~$fornengo@to.infn.it},
C.W. Kim$^{\rm a,c}$
\footnote{E-mail: kim@eta.pha.jhu.edu},
and J. Song$^{\rm a}$
\footnote{E-mail: jhsong@eta.pha.jhu.edu}
}

\vspace{0.8 cm}

\address{
\begin{tabular}{c}
$^{\mbox{a}}$
Department of Physics and Astronomy,
The Johns Hopkins University,\\
Baltimore, Maryland 21218, USA.
\vspace{0.3 cm}
\\
$^{\mbox{b}}$
Dipartimento di Fisica Teorica,
Universit\`a di Torino \\
and Istituto Nazionale di Fisica Nucleare, Sezione di Torino \\
via P. Giuria 1, 10125 Torino, Italy
\vspace{0.3 cm}
\\
$^{\mbox{c}}$
School of Physics,
Korean Institute for Advanced Study
\\
Seoul 130-012, Korea
\end{tabular}
}

\date{February 1997}
\maketitle
\begin{abstract}
$~~$Leading finite temperature effects on 
the neutrino decoupling temperature in the early Universe
have been studied. We have incorporated modifications of the dispersion
relation and the phase space distribution due to the presence of 
particles in the heat bath at temperature
around MeV. Since both the expansion rate of the Universe and the 
interaction rate of a neutrino are reduced by finite temperature 
effects, 
it is necessary to calculate thermal corrections
as precisely as possible in order to find the net effect
on the neutrino decoupling temperature.
We have performed such a calculation by
using the finite temperature field theory. 
It has been shown that the finite temperature effects
increase the neutrino decoupling 
temperature by 4.4\%, the largest contribution 
coming from the modification of the phase space
due to the thermal bath.
\end{abstract}
 
\section{Introduction}

The standard hot big bang model appears to be a reliable description of 
the evolution of the early Universe, one of the most remarkable
successes being the prediction of the present abundance of light
chemical elements from the primordial nucleosynthesis calculations
\cite{crisis}. The
Universe is usually described as a hot, dilute gas of 
particles in nearly thermodynamical 
equilibrium \cite{Kolb,noneq}.
During the early epochs, 
the particle species in the thermal bath 
underwent departure from the equilibrium one after another:
one of the typical departures was that of neutrinos, when
the temperature of the Universe was about 1\,MeV \cite{Kolb}. 
The neutrino decoupling
has important (indirect) effects on the evolution of the Universe, since
it happened at the time close to the 
neutron--to--proton ratio ($n/p$) freeze--out temperature
$T \simeq 0.7$\,MeV and 
to the photon reheating by $e^+ e^-$ annihilation
($T\lsim m_e$). The synthesis of light elements
depends sensitively on the $n/p$ 
freeze--out abundance which is determined
by the interplay between the weak interaction rates
and the expansion rate of the Universe.
Both rates are influenced by the neutrino decoupling 
temperature\cite{Kolb,Dicus}.
Neutrinos which were decoupled early do not share the entropy 
transfer with electrons, positrons and photons in the medium.
As a consequence, their temperature $T_\nu$ becomes slightly lower
than that of the other particles in equilibrium. 
On the contrary,
if neutrinos are not totally decoupled when the entropy transfer begins,
they can share part of the $e^\pm$ entropy and their temperature 
would be higher. A small
change $\Delta T_\nu$ modifies the statistical distribution 
of neutrinos, and in turn affects both the weak interaction rates 
which maintain the equilibrium between neutrons and protons, 
and the expansion rate of the Universe which is due to
the change of the neutrino contribution to the total energy density.
The overall effect is to shift the $n/p$ freeze--out temperature and 
hence the $n/p$ abundance when the nucleosynthesis begins. 
The effect on the present Helium abundance $Y$ 
was estimated to be of the order
$\Delta Y/Y \simeq -0.1 (\Delta T_\nu/T_\nu)$ (the change is 
with respect to
the calculation in which neutrinos are not reheated by 
electrons)\cite{Dicus}.
Therefore, a precise knowledge of the neutrino decoupling temperature
is desirable to gain a confidence in the 
estimates of the primordial element abundance. 

The standard calculation of the neutrino decoupling temperature 
is based on the assumption that 
particles in the thermal bath behave like free
particles. The interactions are only responsible for 
the thermodynamical equilibrium, but do not contribute to the 
energy density of the Universe. However, particles in the medium
feel effective potentials
due to the interactions with other particles,
which modifies their dispersion relations
or introduces effective mass for the particle. 
In addition to this dynamical effect, 
the phase space available for the interaction
is necessarily modified by the statistical distribution
of particles in the medium.

Our purpose is to examine whether or not the finite temperature 
effects can actually lower the neutrino decoupling 
temperature, leading to
possible changes in the nucleosynthesis prediction of the present
abundance of light elements. We explicitly and consistently include
in the calculations the thermal effective mass of a photon and an
electron, and the thermalized phase space distribution
in the cross section in the Born approximation. 
Higher order interactions and
radiative corrections to 
the neutrino interactions 
are not considered here, since we aim
at the leading order corrections by the thermal
bath. Another medium effect, which is not
considered here, is the absorption or the emission 
of photons in the bath. 
This effect turns out to be important in the nucleosynthesis 
calculations for the reactions which involve three body initial or
final state (such as neutron decay and its inverse decay), because the 
additional photon involved modifies sizably
the phase space of the reactions\cite{CPS}.
In the case of two body reactions responsible 
for the thermal equilibrium of neutrinos, however, 
this higher order effect 
is small compared with the corrections 
calculated in the present paper, the
overall magnitude of which will be shown to be 
of the order of 15\% on the
interaction rates, leading to a 4\% shift in the neutrino decoupling
temperature.

The plan of the paper is as follows. In Section 
\ref{sec:thermal}, we discuss how to include finite temperature
effects in the calculation. In particular, we will briefly
discuss the calculation of effective mass of a photon
and an electron in the framework of Finite Temperature Quantum 
Field Theory (FTQFT), with the special attention to the MeV temperature
range which is relevant for the neutrino decoupling. Limitations
of the previous calculations in applying to the present
problem will also be discussed. In Section \ref{sec:decoupling},
we evaluate finite temperature 
effects on the expansion rate of the Universe
and the interaction rate of a neutrino, and show their 
effects on the neutrino decoupling temperature.

\section{Thermal effects}
\label{sec:thermal}

In the early Universe where the particles are propagating in 
a thermal bath, rather than in the vacuum, their dynamics and 
interactions are modified to some extent.
The behavior of particles in a thermal bath is systematically 
described in the framework of FTQFT
\cite{FTQFT}.
There are two equivalent formulations of
FTQFT: the imaginary--time and real--time formalisms.
In the present paper, we adopt the real--time formalism where
the Feynman rules for all the vertices are identical
with those in the vacuum, and the presence of the thermal bath
is taken into account by the modification of the tree--level
propagators of fermions and bosons as \cite{FTQFT}
\begin{equation}
-i S_T(p)	= 
	( \rlap/p + m )\left[
	\frac{1}{p^2-m^2+i\epsilon} 
	+ 2\pi i \delta(p^2-m^2)n_F(p\cdot u)\right]
        \mbox{\hspace{1cm} for fermions}\;,
\label{elprop}
\end{equation}
and
\begin{equation}
- i D^{\mu\nu}_T(k)	=
	\left(-g^{\mu\nu}+\alpha\frac{k^\mu k^\nu}{k^2}\right)\left[
	\frac{1}{k^2+i\epsilon} - 2 \pi i \delta(k^2) \, n_B(p \cdot u)
		   \right]
        \mbox{\hspace{1cm} for photons}\;,
\label{phprop}
\end{equation}
where $\alpha$ is the gauge--fixing parameter.
In Eqs.(\ref{elprop}) and (\ref{phprop}),
$u^{\mu}$ is the four--velocity 
of the medium ($u^{\mu}=(1,\vec{0})$ in the rest frame of the medium)
and $n_{F,B}$ are defined as 
\begin{equation}
n_{F,B}(x)	= \theta(x) f_{F,B}(x) + \theta(-x) f_{F,B} (-x) 
\;,
\label{nfb}
\end{equation}
where $\theta(x)$ is the step function and $f_{F,B}$ are, respectively,
the Fermi--Dirac (FD) and Bose--Einstein (BE) distribution functions 
\begin{equation}
f_{F,B}(x) = 
\left[\exp\left(\frac{x - \eta}{T}\right) \pm  1 \right]^{-1}\;.
\label{distribution}
\end{equation}
In Eq.(\ref{distribution}) the $(+)$ and $(-)$ signs refer to fermions 
and bosons, respectively, and $\eta$ is the chemical potential.
In this paper $\eta = 0$ is assumed for all the species.

Corrections analogous to that of Eq.(\ref{phprop})
should also be applied to the propagators of massive
gauge bosons, which are exchanged in the weak interactions of 
neutrinos. 
Due to the presence of statistical distribution functions,
however,
finite temperature corrections to their vacuum propagators 
are exponentially suppressed at $T \sim 1\,{\rm MeV} \ll M_W,M_Z$. 
This reflects the fact that the bath is too cold to excite those 
very massive degrees of freedom.

In Eq.(\ref{distribution}), $T$ is the temperature of the heath bath
measured in the rest frame of the fluid. The presence of the
bath does not violate the Lorentz invariance of the system 
since
appropriate definitions of temperature, as well as of all
the thermodynamical variables, can be obtained in any reference frame
by suitable transformation laws\cite{Weldon,Tolman}. Since
the early Universe as a thermal bath
is conveniently described in the rest frame of the fluid itself,
i.e. in the comoving reference frame, temperature $T$
has a direct and simple meaning. We will, therefore, adopt the
comoving frame throughout the paper.

In the FTQFT formalism, the effect of the bath on the 
dynamical evolution of particles is taken into 
account by modifications to their dispersion relations, which 
can be recast in the definition of effective mass for the
particle. (This effect can be evaluated by calculating the self--energy
of the particle in the heat bath, and will be briefly 
reviewed in Sections \ref{sec:photon} and \ref{sec:electron}).
The change of mass of the particle modifies its 
contribution to the energy density of the Universe
and therefore its expansion rate.
At the same time, it modifies the interaction 
rate due to the change of dispersion relations in
the cross sections and the distribution functions $f(E)$. 

In addition, the phase space distribution is also influenced by the 
background: the presence of the same particles
in the surrounding medium 
as those produced in the interaction processes reduces (enhances)
the production probability for fermions (bosons), 
respectively, according to the statistics of the particles. 
The final state density factors are
modified as follows:
\begin{eqnarray}
\frac{{\rm d}^3 p\,'}{(2 \pi)^3 2 E'} 
&\longrightarrow&
\frac{{\rm d}^3 p\,'}{(2 \pi)^3 2 E'} [1 - f_F(E')] 
\mbox{\hskip 0.5cm for fermions}
\\
\frac{{\rm d}^3 k\,'}{(2 \pi)^3 2 \omega'}
&\longrightarrow&
\frac{{\rm d}^3 k\,'}{(2 \pi)^3 2 \omega'}  [1 + f_B(\omega')] 
\mbox{\hskip 0.5cm for bosons}
\;.
\end{eqnarray}

In summary, 
the influence of the medium on the particle evolution
in the Universe is due to the temperature--dependent shifts in
the dynamical mass of the particles and the
temperature--dependent modification of the interactions 
between particles.
Both modifications will be included in the calculation of the neutrino
decoupling temperature in Section \ref{sec:decoupling}.
We now turn to the discussion of thermal mass of the photon,
electron and neutrino at $T\sim$ MeV.

\subsection{Effective Mass of the Photon}
\label{sec:photon}

In the thermal bath at $T\sim {\rm MeV}$, a photon propagates through
a medium made of electrons, positrons and neutrinos. Its propagation
is therefore influenced by the interactions with the $e^+$ and $e^-$.
The effect of these interactions on the dynamical evolution of the
photon is taken into account by calculating the self--energy of the
photon in the $e^\pm$ background.
The one--loop self energy diagram gives
\begin{equation}
\Pi^{\mu\nu}(k) = \Pi_0^{\mu\nu}(k) + \Pi_T^{\mu\nu}(k)\;,
\end{equation}
where $k^\mu$ is the photon four momentum, $\Pi_0^{\mu\nu}(k)$ is the
vacuum polarization tensor at $T=0$ and
\begin{eqnarray}
\Pi_T^{\mu\nu} &=& -2\pi e^2 \int \frac{{\rm d}^4 p}{(2\pi)^4}
  	\mbox{Tr} [ \gamma^\mu (\rlap/p+\rlap/k+m_0)\gamma^\nu
  			(\rlap/p+m_0)]
\nonumber \\
  	& & \times \left[\frac{ \delta( p^2-m_0^2)f_F(p^0)}
  		     {(p+k)^2-m_0^2} +
              \frac{ \delta[(p+k)^2-m_0^2]f_F(p^0+k^0)}
  		     {p^2-m_0^2}
        \right]
\;,
\label{eq:pimunupho}
\end{eqnarray}
describes finite temperature corrections.
In Eq.(\ref{eq:pimunupho}), $p^\mu$ is the four--momentum of the
electron in the loop and $m_0$ denotes the electron mass in the vacuum.
The separation of the
self energy into two parts, one referring to the $T=0$ case and
the other coming from the presence of the medium, is attributed 
to the separation of the electron propagator in 
Eq.(\ref{elprop}). The function
$\Pi_0^{\mu\nu}(k)$ is divergent and, as usual, is subject to the
electric charge renormalization. The function $\Pi_T^{\mu\nu}(k)$ 
induces a finite shift in the photon propagator, generating 
effective mass for the photon. 
The vacuum polarization tensor can be decomposed 
as \cite{Weldon,KM}
\begin{equation}
  \Pi^{\mu\nu}=\pi_T(\bar{k},w)\, P^{\mu\nu}
  		+ \pi_L(\bar{k},w) \, Q^{\mu\nu}\;,
\label{eq:pimunu}
\end{equation}
where $P^{\mu\nu}$ and $Q^{\mu\nu}$ are orthogonal projection operators
(for their explicit form, see \cite{Weldon}) and $\pi_T$ and 
$\pi_L$ are scalar functions given by
\begin{eqnarray}
  \pi_L(\bar{k},w) 
  &=& -\frac{k^2}{\bar{k}^2}
  		u^\mu u^\nu \Pi_{\mu\nu}
\nonumber  \\ 
  \pi_T(\bar{k},w) &=& -\frac{1}{2} \pi_L(\bar{k},w)
  	+ \frac{1}{2} g^{\mu\nu} \Pi_{\mu\nu} \;.
\label{eq:pi}
\end{eqnarray}
In Eqs.(\ref{eq:pimunu}) and (\ref{eq:pi}), $w = k^\mu u_\mu$
and $\bar{k} = \sqrt{w^2 - k^\mu k_\mu}$. In the comoving frame,
$w(=k^0)$ is the energy of the photon and $\bar{k} (= |\vec k|)$
denotes the magnitude of its three--momentum. The decomposition in 
Eq.(\ref{eq:pimunu}) allows one to write the propagator 
as (in the Feynman gauge)
\begin{equation}
  \Delta^{\mu\nu}=-\frac{P^{\mu\nu}}{k^2-\pi_T} -\frac{Q^{\mu\nu}}
  		{k^2-\pi_L}\;.
\label{eq:deltamunu}
\end{equation}
In the comoving frame,
the $P^{\mu\nu}$ term describes the transverse modes
of the photon field, while the $Q^{\mu\nu}$ term 
is a linear combination of longitudinal
and time--like modes. Equation (\ref{eq:deltamunu}) shows that the
propagator of physical (transverse) photon modes has a pole at
$k^2 = \pi_T$. 
This is interpreted as a thermal--generated dynamical mass. 
The pole $k^2 = \pi_L$ in the non--transverse part of the propagator
describes the Debye screening length of the photon 
in the $e^\pm$ plasma\cite{Weldon,Masood}.

Since we are interested in the leading order $O(\alpha)$
correction to the mass of the photon, the vacuum dispersion relations 
$k^2=0$ for the photon and $p^2=m_0^2$ for the electron can be used
in the right hand sides in Eqs.(\ref{eq:pimunu}) and
(\ref{eq:pi}).
Therefore, the effective mass of the photon is
\begin{equation}
  [m_\gamma^{\rm eff}(T)]^2 = {\rm Re}\left[\pi_T(\bar{k},w)
\right]=
   \frac{8\alpha}{\pi} T^2 h(\mu_0)
+{\cal O\/} (\alpha^2)
\;,
\label{eq:mphoton}
\end{equation}
where $\mu_0 = m_0/T$ and the function $h(\mu_0)$ is defined as 
\begin{equation}
  h(\mu_0) =\int_0^{\infty} 
{\rm d}x
\,\frac{x^2\,
}{\sqrt{x^2+(\mu_0)^2}}
\,f_F(x)\;.
\label{eq:funta}
\end{equation}
Note that the photon effective mass in Eq.(\ref{eq:mphoton})
has been obtained without any restriction to $T$, hence valid
for all temperature. Equation (\ref{eq:mphoton})
gives the correct limit $m_\gamma=0$ for $T=0$ and is in agreement
with the
result $[m_\gamma^{\rm eff}(T)] = (2\pi\alpha T^2)/3$ obtained in
Ref.\cite{Weldon} in the limit of high temperature $T\gg w, m_0$.
Obviously this limit does not apply to the early Universe 
where photons are in thermal equilibrium with 
the mean energy $w \sim T$. 
Moreover, since $T \sim 
m_0
$ at the neutrino decoupling temperature,
the limit $
\mu_0/x=m_0/T
 \ll 1$ in Ref.\cite{Masood} cannot be applied.

The thermal mass of the photon is almost linearly dependent on 
temperature. 
The photon effective mass is 0.115 MeV at $T=1$ MeV  and
0.241 MeV at $T=2$ MeV.
Even though the photon is still relativistic, 
its contribution to the energy density of the Universe is 
substantially reduced.

\subsection{Effective Mass of the Electron}
\label{sec:electron}

As in the case of the photon, the dynamics of electrons 
in a thermal bath is also
modified by the electromagnetic interactions with background photons
and electrons themselves. The interactions
with neutrinos are suppressed at $T\sim 1$ MeV, for they involve
the exchange of heavy bosons $W$ and $Z$. Therefore, 
the effect of the 
bath on the propagation of the electron is expressed by calculating
the electron self--energy in the presence of the ambient $e^+$, $e^-$
and $\gamma$.
The electron self--energy at one--loop level becomes
\begin{equation}
\Sigma(\rlap/p) = \Sigma_0(\rlap/p) + \Sigma_T(\rlap/p)
\;,
\end{equation}
where $\Sigma_0(\rlap/p)$ is the electron self--energy 
for $T=0$ and its thermal correction is
\begin{eqnarray}
\Sigma_T(\rlap/p) &=& -2\pi e^2 \int \frac{{\rm d}^4 k}{(2\pi)^4}
 [\gamma^\mu(\rlap/p+\rlap/k+m_0)\gamma_\mu]
\label{eq:sigmaT}
\\ \nonumber
& & \times
    \left[
     \frac{\delta(k^2) f_B(k^0)}{(p+k)^2 - m_0^2} -
     \frac{\delta[((p+k)^2-m_0^2) f_B(p^0+k^0)}{k^2} 
    \right]
\end{eqnarray}
where $p^\mu$ is the four--momentum of the electron and $k^\mu$ denotes
the momentum in the electron--photon loop.
The one--loop self--energy modifies the electron propagator $S(p)$ as
\begin{equation}
S(p)^{-1} = \rlap/p - m_0 - \Sigma(\rlap/p) =
  [\rlap/p - m_0 - \Sigma_0(\rlap/p)] - \Sigma_T(\rlap/p)
\;.
\end{equation}
The standard $\Sigma_0(\rlap/p)$ leads to the definition 
of physical mass of electrons and the
wave function renormalization in the vacuum. The temperature--dependent
self--energy $\Sigma_T(\rlap/p)$ produces a finite shift 
in the dispersion relation as 
\cite{Donoghue}
\begin{eqnarray}
  [m_e^{\rm eff}(T,p)]^2 
&\equiv& E^2-\vec{p}\,^2
\label{eq:elmeff}
\\ \nonumber
&=&
 m_0^2 + \frac{2}{3}\alpha\pi T^2
  			+ \frac{4\alpha^2}{\pi} h(\mu_0)
  			+\frac{\alpha}{2\pi^2}m_0^2J_A(p)\;,
\end{eqnarray}
where the function $h(\mu_0)$ is defined in Eq.(\ref{eq:funta}) and
\begin{equation}
J_A(p) = -\frac{2\pi}{u}
\int_0^{\infty} \,\frac{x\,{\rm d}x}{\sqrt{x^2+(\mu_0)^2}} \; 
f_F(\epsilon_x)\; \ln \left(
 \frac{\epsilon_p \epsilon_x + \mu_0^2 + xu}
      {\epsilon_p \epsilon_x + \mu_0^2 - xu} \;\;
 \frac{\epsilon_p \epsilon_x - \mu_0^2 + xu}
      {\epsilon_p \epsilon_x - \mu_0^2 - xu} \right)\;,
\end{equation}
where dimensionless quantities have been defined as $u=|\vec p|/T$,
$\epsilon_p = E_p/T$, $x=|\vec p - \vec k|/T$ and 
$\epsilon_x = (E_p-E_k)/T$. The result of Eq.(\ref{eq:elmeff}) is valid
for all temperature. It gives the correct result
$m_e^{\rm eff} = m_0$ at $T=0$ and agrees with the result of
\cite{Donoghue,Ahmed,Weldon-p} 
in the limiting cases $T\ll m_0$ and $T\gg m_0$. The last two
terms in Eq.(\ref{eq:elmeff}) are relatively small at $T\ll m_0$, 
for they are exponentially suppressed 
by the Fermi--Dirac or the Bose--Einstein distribution
function. Around $T \sim m_0$, however,
the third term becomes important and has to be taken into
account (for instance, $h(m_0/T=1) \simeq 0.543$).
The last term $J_A(p)$ in Eq.(\ref{eq:elmeff}) introduces momentum
dependence in the electron effective mass, which is important
especially in the calculation of the neutrino cross sections 
with the momentum--dependent phase space distributions. 
This term has been calculated 
in the limit $\epsilon_x\gg \mu_0$\cite{Ahmed}. 
This approximation is only valid at high temperature
(T$\gg m_0$), 
for the most significant contribution of $\epsilon_x$ to the
integral is from its mean values, i.e. $\epsilon_x \sim 1$. 
In the temperature range of our
interest ($T\sim m_0$), this approximation
cannot be adopted and we have to resort to a numerical calculation.
This function is always negative, and
monotonically increasing from the limiting value
\begin{equation}
  \lim_{p\rightarrow 0} J_A(p) = -8\pi
\int_0^{\infty} \,\frac{{\rm d}x}{\sqrt{x^2+(\mu_0)^2}} \, 
f_F(\sqrt{x^2+\mu_0^2})
 \;,
  \end{equation}
up to $\lim_{p\to\infty}J_A(p) = 0$.
In order to see how significant the modifications to $m_0$
due to the $J_A(p)$ term is,
we plot in Fig.1 the deviation of the $m_e^{\rm eff}$ 
calculation including the $J_A(p)$ term from that neglecting $J_A(p)$,
as a function of temperature.
The dashed line refers to the calculation for the case with
$p=\langle p \rangle$, 
where $\langle p\rangle$ is the mean value of the momentum in the bath. 
The solid line referring to the case with $p=0$ 
corresponds to the largest contribution of the $J_A(p)$ term,
the maximum of which is $3.3 \times 10^{-3}$ at $T \sim 2$~MeV. 
The other thermal corrections in Eq.(\ref{eq:elmeff}) 
are always larger than that due to the
$J_A(p)$ term at least by one order of magnitude, 
as can be seen by comparing Fig.1 to Fig.2 (in Fig.2, the relative
correction $|1-m_e^{\rm eff}(T)/m_0|$ is plotted as a function of $T$). 
This shows that the $J_A(p)$ term
is negligible around the temperature $T\sim$ MeV. We will therefore
neglect this term in our analysis.
This turns out to be a great simplification in the calculations,
especially for the interaction rate, because the momentum dependence
in the electron mass is avoided.
  
Figure 2 shows that the thermal corrections to the electron mass
at $T\sim $ MeV is sizeable: at $T=1$ MeV
the electron mass increases by 4.1\%, and at $T=2$ MeV the correction
is as large as 16\%.

\subsection{Effective mass of the neutrino}
  
Neutrinos also acquire effective mass in the presence of a
medium. In the temperature range of our interest,
the contributions come from weak interactions with the ambient
electrons and positrons. The effective mass squared of the neutrino
has been shown to be of the order of $G_F (N-\overline{N})$,
where $G_F$ is the Fermi constant and 
$N(\overline{N})$ is the number density 
of electrons(positrons)\cite{Kim}. Since we expect
$N \simeq \overline{N}$, the modification 
to the neutrino effective mass at $T\sim 1$ MeV is practically absent.
Therefore, we neglect finite temperature effects
on the dispersion relation of neutrinos in the following.

\section{Neutrino decoupling}
\label{sec:decoupling}

Neutrinos in the early Universe, like all other particles,
are kept in thermodynamical equilibrium
through their interactions with the particles in the
heat bath. As long as its interaction rate $\Gamma$ is larger than 
the expansion rate of the Universe $H$, 
neutrinos remain in thermal and chemical equilibrium. 
When $\Gamma$ becomes smaller than $H$, due to 
the reduced temperature of the heat bath and therefore
the increased distance between particles,
neutrinos start to depart from the equilibrium and subsequently
evolve independently from the other species. 
Even though this decoupling is not a sharp event, 
we can define that it happens when 
\begin{equation}
\Gamma(T_d) = H(T_d)\;.
\label{decoup}
\end{equation}
The temperature when Eq.(\ref{decoup}) holds is defined as the
neutrino decoupling temperature $T_d$. 
In the following Subsections we will
discuss finite temperature effects on the
calculation of $T_d$. Both $\Gamma$ and $H$ decrease as 
finite temperature effects are incorporated. 
A detailed calculation is therefore needed
to determine whether or not the decoupling temperature actually 
increases or decreases, and to estimate the size of the effect.

\subsection{Expansion rate}

In the standard big bang model,
the dynamical expansion of the early Universe is governed by the
Friedman equation \cite{Kolb}
\begin{equation}
H = \left [ \frac{8 \pi G}{3} \rho \right ] ^{1/2}\;,
\label{friedman}
\end{equation}
where $G$ is the Newton constant.
Under the assumption that all the particles are in 
thermodynamical equilibrium, the total energy density $\rho$ is
\begin{equation}
\rho = \sum_i g_i \int \frac{{\rm d}^3 p}{2\pi^3}\; E(p) \, f_i(E) \;,
\label{rho}
\end{equation}
where $g_i$ is the number of internal degrees of freedom of the
particle species $i$. 
$E$ and $p$ are its energy and momentum,
respectively, which are related by the usual dispersion
relation $E^2 = \vec{p}^2 + m_i^2$,
where $m_i$ is the mass of the particle.

Equation (\ref{rho}) leads to the following expression of $H$:
\begin{equation}
H = \sqrt{\frac{4 \pi^3 G}{45}} g^{1/2}_*(T) T^2\;,
\label{hubble}
\end{equation}
where $g_*(T)$ is the number of degrees of freedom at 
temperature $T$, defined by
\begin{equation}
g_*(T) = \sum_i \left( \frac{T_i}{T} \right)^4 \frac{15 g_i}{\pi^4}
\int_0^\infty 
{\rm d}u\; 
\frac{
u^2 \sqrt{u^2 + \mu_i^2}}
      {\exp\left(\sqrt{u^2 + \mu_i^2}\right) \pm 1}\;,
\label{dof}
\end{equation}
where $\mu_i = m_i/T$. 
In Eq.(\ref{dof}),
$T_i$ is the temperature of the species $i$. 
For the species $i$ in equilibrium,
$T_i$ is equal to the temperature $T$ of the Universe, but
after the decoupling, its temperature does not need to 
be the same as $T$ (this is actually the case for neutrinos).
Note that the dimensionless $g_*(T)$ is proportional to the
energy density of the Universe in units of $T^4$, which is
effectively dominated by highly relativistic
particles; since particles near the transition from the relativistic to
the non--relativistic regime can also contribute to $g_*(T)$,
they should be also taken into account in a precise calculation. 
When a particle species becomes non--relativistic, 
its contribution to the energy density and therefore to $g_*(T)$
is exponentially suppressed.
Around the neutrino decoupling temperature ($T\simeq 1$\,MeV), 
the particles in thermal equilibrium are photons, 
electrons, positrons and $\nu_e(\bar{\nu}_e)$. 
We will assume that muon and tau neutrinos had already been decoupled, 
so that they do not interact with the $\nu_e$. 
However, they still contribute to the energy density.

Now, let us discuss the effects of the thermal background.
Since electrons, positrons and photons have effective mass larger
than  that in the $T=0$ situation,
$g_*(T)$ and hence $H(T)$ are reduced. 
Figure 3 shows the relative change $[1-g_*^{\rm FT}(T)/g_*(T)]$ as a
function of temperature. 
Although the photon mass is a few tenth of MeV and the electron
mass increases by $5 \sim 10$\% at the temperatures 
around $1 \sim 2$ MeV, 
the effect on the number of degrees of freedom is small:
$g_*(T)$ decreases by a factor of 0.1\%.
This is because in this temperature regime 
$g_*(T)$ is dominated by three neutrino species. Due to
the relation between $H(T)$ and $g_*(T)$ 
in Eq.(\ref{hubble}), the ensuing effect
on the expansion rate is a factor of two smaller:
$\Delta H/H \simeq 0.5 (\Delta g_*/g_*) \simeq 5\times 10^{-4}$.
As will be shown in Section \ref{sec:interaction}, 
finite temperature modifications to $\Gamma$ are 
two order of magnitude larger than that to the expansion
rate, leaving very little influence of $H$ on the
calculation of $T_d$.

\subsection{Interaction rate and neutrino decoupling temperature}
\label{sec:interaction}

Neutrinos are kept in thermal equilibrium by the interactions with 
electrons and positrons in the heat bath. 
Focusing on the electron neutrino 
$\nu_e$, the interactions such as
\begin{eqnarray}
\nu_e + e^- &\longleftrightarrow& \nu_e + e^- \label{eq:nue}\\
\nu_e + e^+ &\longleftrightarrow& \nu_e + e^+ \label{eq:nue2} \;,
\end{eqnarray}
are responsible for kinetic equilibrium, and annihilation and
creation processes like
\begin{equation}
\nu_e + \bar\nu_e \longleftrightarrow e^- + e^+  \;,
\label{eq:nue3}
\end{equation}
maintain neutrinos in chemical equilibrium.
Since the interactions which involve more 
than two particles 
are suppressed by additional powers of the small coupling constants,
we will not consider them in the following.
For definiteness, we will discuss
in detail the interaction rate for the process of 
Eq.(\ref{eq:nue}), which has the largest cross section and therefore is
dominant in the determination of the thermal equilibrium for
$\nu_e$. We will turn to the other processes at the end
of this Section.

The standard calculation of the interaction rate of a neutrino relies 
on a number of simplifying assumptions: (1)
the electron is considered to be massless, i.e. $m_e=0$; 
(2) the energy distribution of the initial--state particles is
neglected: two initial particles are considered to have mean--valued
energies
\begin{eqnarray}
\langle E \rangle &=& \frac{g_e}{N_e(T)}\int\frac{{\rm d}^3 p}{(2\pi)^3}
\,E\,f_e(E) \\
\langle \omega \rangle &=& \frac{g_\nu}{N_\nu(T)}\int
\frac{{\rm d}^3 k}{(2\pi)^3}\,\omega\,f_\nu(\omega) \;,
\end{eqnarray}
and the interaction is supposed to occur in the center of momentum frame
($\cos\theta=-1$).
Under these assumptions, the interaction rate becomes
\begin{equation}
\Gamma = N_e(T)\,v_M\,
\sigma(\langle E \rangle,\langle \omega \rangle) \;,
\end{equation}
where
 $v_M = 2$.
The cross section $\sigma$ for the massless electron is
\begin{eqnarray}
\sigma = \frac{G_F^2 A}{12\pi}\,s \;,
\end{eqnarray}
where
$s=4 \langle E \rangle\langle \omega \rangle$
and the definition of the constant $A$ will be given later
(see Eq.(\ref{def-A})).
Since the electron
is assumed to be massless, we have
\begin{equation}
\langle E \rangle=\langle \omega \rangle= \frac{7\pi^4}{180\zeta(3)}\;T 
\simeq 3.15 \;T
\end{equation}
and
\begin{equation}
N_e(T) = \frac{3}{4}\frac{\zeta(3)}{\pi^2}g_e T^3 \simeq 0.182\, T^3 \;,
\end{equation}
where $\zeta$ is the Riemann zeta function and 
$\zeta(3) = 1.20$.
Consequently,
$\Gamma$ is
\begin{equation}
\Gamma \simeq 0.385 \,A G_F^2 T^5 \simeq 3.48\times 10^{-22} 
\left(\frac{T}{{\rm MeV}}\right)^5 \;.
\label{eq:gamma-rough}
\end{equation}
Comparing Eq.(\ref{eq:gamma-rough}) with the expansion rate calculated
for the massless electron ($g_*(T) = 10.75$),
\begin{equation}
H \simeq 4.46\times 10^{-22} 
\left(\frac{T}{{\rm MeV}}\right)^2 
\,,
\end{equation}
the decoupling temperature is
\begin{equation}
T_d \simeq 1.09\; {\rm MeV} \;.
\label{eq:tdec-rough}
\end{equation}
The above $standard$ analysis provides very rough
estimates of the interaction rate and the decoupling temperature.
For comparison, the calculation of the expansion rate for $m_e=0$,
but taking into account the thermal 
distribution of the initial energies, gives
\begin{equation}
\Gamma \simeq 0.1284 \,A G_F^2 T^5  \;,
\label{eq:gamma-better}
\end{equation}
which is a factor of 3 smaller than the estimate in 
Eq.(\ref{eq:gamma-rough}). The resulting decoupling temperature is
\begin{equation}
T_d \simeq 1.56\; {\rm MeV} \;,
\end{equation}
which is 50\% higher than the value given in Eq.(\ref{eq:tdec-rough}).

A detailed and precise calculation of the interaction rate is rather
involved, but necessary to see whether or not finite temperature 
effects lower the neutrino decoupling temperature.
For the following interaction
\begin{equation}
\nu_e(\omega, \vec{k}) + e^- (E,\vec{p}) 
\longleftrightarrow \nu_e(\omega', \vec{k}') + e^- (E',\vec{p}') \;,
\label{process}
\end{equation}
the interaction rate is defined by
\begin{equation}
\Gamma (\nu_e e^- \rightarrow \nu_e e^-) = \frac{1}{N_\nu(T)}
\int \frac{{\rm d}^3 p}{(2\pi)^3} \frac{{\rm d}^3 k}{(2\pi)^3}
\, g_e f_e(E)\, g_\nu f_\nu(\omega) \; [\sigma v_M] \;,
\label{eq:gamma}
\end{equation}
where the number density $N_\nu(T)$ of neutrinos at temperature $T$ is
\begin{equation}
N_\nu(T) = g_\nu \int \frac{{\rm d}^3 k}{(2 \pi)^3} f_\nu(\omega) \;.
\end{equation}
That is, $\Gamma$ is the thermal average of 
product of the cross section 
$\sigma$ and the M\"oller velocity $v_M$ \cite{Gelmini}. 
The cross section of the process is, including the thermal phase space,
\begin{equation}
\sigma = \frac{1}{4 E \omega v_M} 
\int \frac{{\rm d}^3 p'}{(2\pi)^3\,2E'} 
     \frac{{\rm d}^3 k'}{(2\pi)^3\,2\omega'}
     [1-f_e(E')]\,[1-f_\nu(\omega')] \;
     (2\pi)^4 \delta^{(4)}(p+k-p'-k') |{\cal M}|^2 \;,
\label{eq:sigma-full}
\end{equation}
and the M\"oller velocity is defined (for massless neutrinos) by
\begin{equation}
v_M = \frac{p^\alpha k_\alpha}{E\omega} = 
\frac{s}{2E\omega} \left( 1-\frac{m_e^2}{s}\right) \;,
\label{eq:vmoller}
\end{equation}
where $s \equiv (p+k)^\alpha (p+k)_\alpha$ 
is the total energy in the center of momentum frame of the
colliding particles and its expression in the comoving frame is
($\theta$ is the angle between 
$\vec p$ and $\vec k$)
\begin{equation}
s = m_e^2 + 2E\omega -2pk\cos\theta \;.
\end{equation}

The scattering amplitude for the process 
$\nu_e e^- \longrightarrow \nu_e e^-$
comes from two diagrams with $Z$  in the $t$--channel and 
$W$  in the $u$--channel. Since mean energies of interacting
particles are of the order of the temperature 
$T\simeq$\,MeV$(\ll M_{W,Z})$, we can
express the averaged square amplitude in the low energy limit as 
\begin{equation}
|{\cal M}|^2 = 16 G_F^2 
[ (v+a)^2 (p^\alpha k_\alpha) (p'^\alpha k'_\alpha) +
  (v-a)^2 (p'^\alpha k_\alpha) (p^\alpha k'_\alpha) -
  (v^2-a^2)^2 m_e^2 (k^\alpha k'_\alpha) ] \;,
\end{equation}
where $G_F$ is the Fermi constant, $v=g_V+1$, $a=g_A+1$, and $g_V$
($g_A$) are the vector (axial--vector) weak coupling constant.

As a first step, 
we neglect the thermal phase space in order to investigate
the effect of electron thermal mass alone. 
With the standard phase space, the cross section is 
\begin{equation}
\sigma = \frac{G_F^2}{12\pi}\,s\,\left(1-\frac{m_e^2}{s}\right)^2
\,\left[A + B \frac{m_e^2}{s} + C \frac{m_e^4}{s^2}\right] \;,
\label{eq:sigma}
\end{equation}
where
\begin{eqnarray}
A &=& 4(a^2+av+v^2) 
\label{def-A}
\\
B &=& 2(2 a^2-av-v^2) \\
C &=& (a-v)^2 \;.
\end{eqnarray}
Finally, $\Gamma$ is to be obtained by taking a thermal
average of $\sigma v_M$.
A simple expression of $\Gamma$ in terms of Bessel functions, 
involving a one--dimensional numerical integration, 
has been obtained in the case 
of Maxwell--Boltzmann (MB) distribution for the initial 
particles\cite{Gelmini}. 
We do not adopt this approximation here, 
because it induces larger 
uncertainties than the finite temperature corrections 
in the present analysis. 
For instance,
particles with energy $E\sim \langle E \rangle$ would have
4.7\% larger interaction rate in the MB distribution than
in the FD distribution.
Therefore, we have to resort to an improved method\cite{DK}.

Since the only angular dependence comes 
from the relative angle $\theta$ 
between $\vec p$ and $\vec k$, we have
\begin{equation}
{\rm d}^3 p\;{\rm d}^3 k = 
4\pi p^2{\rm d} p \; 2\pi k^2{\rm d} k \; {\rm d}\!\cos\theta \;,
\end{equation}
with the kinematical limits $(0,\infty)$ for both $p$ and $k$ and
$|\cos\theta| \le 1$. With the isotropic distribution functions,
the annihilation rate is
\begin{equation}
\Gamma = \frac{1}{N_\nu(T)} \frac{g_e g_\nu}{8 \pi^4} \frac{G_F^2}{12\pi}
\int_0^\infty \frac{p^2 {\rm d}p}{E} f_e(E) \;
\int_0^\infty \frac{k^2 {\rm d}k}{\omega} f_\nu(\omega) \; 
I(p,k) \;,
\label{eq:gamma2}
\end{equation}
where 
\begin{eqnarray}
I(p,k) &=& \frac{1}{2} \int_{-1}^{1} {\rm d}\cos\theta
\;s^2\,\left(1-\frac{m_e^2}{s}\right)^3 \nonumber
\,\left[A + B \frac{m_e^2}{s} + C \frac{m_e^4}{s^2}\right] \\
&=& \frac{1}{2} \sum_{i=1}^6 a_i I_i \;,
\end{eqnarray}
where the constants $a_i$ and the functions $I_i$ are obtained
from the angular integration as
\begin{eqnarray*}
a_1    &=& m_e^4 (C+3A-3B) \\
a_2    &=& m_e^2 (B-3A) \\
a_3    &=& A \\
a_4 &=& m_e^6 (3B-A-3C) \\
a_5 &=& m_e^8 (3C-B) \\
a_6 &=& -m_e^{10} C \;,
\end{eqnarray*}
and
\begin{eqnarray*}
I_1    &=& 2 \\
I_2    &=& 2 (m_e^2+2E\omega) \\
I_3    &=& 2 (m_e^2+2E\omega)^2 + \frac{8}{3}p^2\omega^2 \\
I_4 &=& \frac{1}{2p\omega} 
 \ln\left[\frac{
m_e^2+2E\omega+2p\omega
}{
m_e^2+2E\omega-2p\omega
}
\right] \\
I_5 &=& \frac{1}{2p\omega} 
 \left[\frac{1}{m_e^2+2E\omega-2p\omega} -
  \frac{1}{m_e^2+2E\omega+2p\omega}\right] \\
I_6 &=& \frac{1}{4p\omega} 
 \left[\frac{1}{(m_e^2+2E\omega-2p\omega)^2} - 
  \frac{1}{(m_e^2+2E\omega+2p\omega)^2}\right] \;.
\end{eqnarray*}
The above holds for all interaction rate 
as long as the involved electron has
momentum--independent mass.
The integrations over $p$ and $k$ in Eq.(\ref{eq:gamma2}) are performed
numerically (all the numerical calculations are performed 
with a precision of $10^{-5}$ but only three significant decimals
are shown).

The interaction rate for $m_e = m_0$,
i.e. without any thermal correction, is plotted in Fig.4 (dotted line)
as a function of temperature. 
The decoupling condition Eq.(\ref{decoup})
is satisfied at the temperature
\begin{equation}
T_d (m_e=m_0) = 1.57\; {\rm MeV}\;.
\end{equation}
The inclusion of the electron thermal mass
has the effect to slightly
reduce the interaction rate
(see the dashed line in Fig.4): 
around $T\simeq 1.6$\,MeV the reduction is 0.3\%.
As a consequence, the decoupling temperature increases by 0.1\%, i.e.
\begin{equation}
T_d (m_e=m_e^{\rm eff}) = 1.58\; {\rm MeV}\;.
\end{equation}

Let us now take into account the
thermal phase space in the evaluation of the cross section, which is
\begin{equation}
{\rm d} \Pi = \frac{{\rm d}^3 p'}{(2\pi)^3 2E'}
  \frac{{\rm d}^3 k'}{(2\pi)^3 2\omega'} [1-f_e(E')] [1-f_\nu(\omega')]
  (2\pi)^4 \delta^{(4)}(p+k-p'-k') \;.
\end{equation}
The three--momentum $\vec p$ can be integrated out by using 
the three--dimensional delta function to give
\begin{equation}
{\rm d} \Pi = \frac{1}{(2\pi)^2 4E'\omega'}
  k'^2{\rm d}k' {\rm d}\phi_{k'} {\rm d}\!\cos\theta_{k'}
  [1-f_e(E')] [1-f_\nu(\omega')]
  \delta(E+\omega-E'-\omega') \;,
\end{equation}
where the angle $\theta_{k'}$ is defined as
\begin{equation}
\cos\theta_{k'} =
\frac{(\vec p + \vec k) \cdot \vec k'}{|\vec p + \vec k| |\vec{k}'|} \;.
\end{equation}
After a trivial integration over $\phi_{k'}$,
we use the remaining one--dimensional delta function to perform the
integration over $\cos\theta_{k'}$
\begin{equation}
{\rm d}\Pi = \frac{1}{8\pi} \frac{\omega'}{E'} 
  \left| \frac{\partial(E'+\omega')}{\partial\cos\theta_{k'}} \right|^{-1}
  [1-f_e(E')] [1-f_\nu(\omega')] {\rm d}\omega' \;,
\end{equation}
with the constraint $\vec p + \vec k = \vec p\,' + \vec k\,'$. This gives
\begin{equation}
{\rm d}\Pi = \frac{1}{8\pi {\cal A}} P(\omega'){\rm d}\omega' \;,
\end{equation}
where we have defined
\begin{equation}
P(\omega') = [1-f_e(E'(\omega'))] [1-f_\nu(\omega')] \;,
\end{equation}
with
\begin{eqnarray}
E'(\omega') &=& \alpha - \omega' \\
\alpha      &=& E + \omega \\
{\cal A}    &=& |\vec p + \vec k| = (p^2 + k^2 +2 p k \cos\theta)^{1/2} \;.
\end{eqnarray}
The quantities $\alpha$ and ${\cal A}$ are fixed 
with the given initial state variables. 
The kinematical limits for the final state energy $\omega'$ are
\begin{eqnarray}
\omega'_{\rm max} &=& 
\frac{\alpha^2 - {\cal A}^2 - m_e^2}{2(\alpha-{\cal A})}
 \\
\omega'_{\rm min} &=& 
\frac{\alpha^2 - {\cal A}^2 - m_e^2}{2(\alpha+{\cal A})} 
    \;,
\end{eqnarray}
which can be obtained from the constraint $|\cos\theta_{k'}| \leq 1$.

The relevant phase--space distribution $P(\omega')$ is a function of the
energy $\omega'$ and of the energies and momenta of initial particles. 
In the case of non--thermal phase space,  we have
$P(\omega') = 1$. The function
$P(\omega')$ represents the reduction of the phase--space due to 
thermal effects. 
Figure 5 shows
as a function of $\omega'$
 $P(\omega',E=\langle E \rangle, \omega
=\langle \omega
 \rangle)$
with $\cos\theta = 0$ 
at three different temperature $T=0.1$, 1, 5 MeV.
We observe that the reduction due to the thermal distribution
function is sizeable over all the energy range of $\omega'$. 
Especially around the temperature $T \sim$ MeV, 
the amount of the reduction is always
greater than 8\% for all energies.

Unfortunately, the functional form of $P(\omega')$ prevents us from 
analytically performing the integration over the final state energy 
$\omega'$ and the initial state angle $\theta$, contrary to the
previous situation without the thermal phase space.
In order to minimize numerical integrations, we 
approximate the effect of the thermal phase--space as a mean value
effect
\begin{eqnarray}
4E\omega (\sigma v_M)_{TH} &=& 
    \frac{1}{8\pi{\cal A}} \int P(\omega'){\rm d}\omega' |{\cal M}^2| 
\nonumber \\
    &\simeq& \frac{1}{8\pi{\cal A}} \langle P(\omega') \rangle 
        \int {\rm d}\omega' |{\cal M}^2| = 
  4 E\omega (\sigma v_M)_{0} \langle P(\omega') \rangle \;,
\label{eq:erre}
\end{eqnarray}
where the quantity $E\omega (\sigma v_M)_{0}$ is
the same as the one calculated previously (see Eqs.(\ref{eq:vmoller}) 
and (\ref{eq:sigma})). In Eq.(\ref{eq:erre}) 
the mean value of the phase space is defined as
\begin{equation}
\langle P(\omega') \rangle \equiv R(T;E,\omega)=
    \frac{1}{2} \int_{-1}^1 {\rm d}\cos\theta 
    \frac{1}{\omega'_{\rm max} - \omega'_{\rm min}} 
        \int_{\omega'_{\rm max}}^{\omega'_{\rm min}} P(\omega'){\rm d}\omega'
    \;.
\end{equation}
The reduction factor $R(T;E,\omega)$ is a function of the initial state 
energies and the temperature. In Fig.6, $R(T;E,\omega)$ is plotted
against the temperature for
$E=\langle E \rangle$ and $\omega=\langle \omega \rangle$. 
As the temperature increases, $R$ decreases
because the Fermi blocking
at higher temperature obstructs scattering processes.
For instance, $R(\langle E \rangle,\langle \omega \rangle) = 0.86$
at $T=1$\, MeV. Note that the effect even at $T=0.1$\, MeV
is about 8\% increase, where the finite temperature effects due to
the thermal mass are totally negligible.

With the definition of the reduction factor $R(T;E,\omega)$,
the interaction rate including all the thermal effects is
\begin{equation}
\Gamma = \frac{1}{N_\nu(T)} \frac{g_e g_\nu}{8 \pi^4} 
\frac{G_F^2}{12\pi}
\int_0^\infty \frac{p^2 {\rm d}p}{E} f_e(E) \;
\int_0^\infty \frac{k^2 {\rm d}k}{\omega} f_\nu(\omega) \; 
I(p,k) R(T;E,\omega) \;.
\label{eq:gamma-final}
\end{equation}
The interaction rate of Eq.(\ref{eq:gamma-final}) is plotted in Fig.4
as a solid line. The global reduction of $\Gamma$, compared to the 
situation without the thermal phase space, is 13\%
for $T\sim$ MeV.
This is by far the most important effect due to the presence of the
thermal bath, being more than a factor of 40 larger 
than the inclusion of the electron thermal mass 
alone in the calculation of $\Gamma$. We 
recall that the thermal mass affects the expansion rate much less,
leading to $\Delta H/H \sim 5\times 10^{-4}$ 
in the MeV range of the temperature. 
The decoupling temperature obtained from $\Gamma$ in 
Eq.(\ref{eq:gamma-final}) is, finally,
\begin{equation}
T_d (m_e = m_e^{\rm eff}, 
{\rm thermal\, phase\, space}) = 1.643\; {\rm MeV} \;,
\end{equation}
which is a 4.4\% increase from the decoupling temperature
without any thermal effect.
We, therefore, conclude that the thermal bath has 
the effect to increase the neutrino decoupling temperature,
assuring that
neutrinos are totally decoupled 
at the time of $e^\pm$ entropy release.

For completeness, we conclude this Section by reporting 
the neutrino decoupling temperature 
taking into account all the
reactions listed in Eqs.(\ref{eq:nue}--\ref{eq:nue3}). Including the
thermal mass and the thermal phase space as
discussed above, the neutrino decoupling temperature increases by
4.4\% into
\begin{equation}
T_d \simeq 1.41\; {\rm MeV} \;,
\end{equation}
as compared to $T_d \simeq 1.35$ MeV obtained
without finite temperature effects.

\section{Conclusions}
We have studied leading finite temperature effects on the neutrino 
decoupling temperature 
in the early Universe. The major motivation
is to investigate if finite temperature effects could actually lower 
$T_d$ and eventually affect the nucleosynthesis calculations.
Two major features of the finite temperature effects  
have been incorporated in the calculation:
(1) the interactions among the particles in the medium affect their
dispersion relations which are recast in the form of
effective mass ;
(2) the presence of the medium modifies the phase space
distribution of the particles in the processes that determine 
the equilibrium.

The effect of the inclusion of thermal mass into the expansion
rate of the Universe $H$ has been shown very small:
$\Delta H/H \simeq 5 \times 10^{-4}$. This effect turns out to
be negligible compared to the modification of
the interaction rate of a neutrino due to
finite temperature. The latter has been discussed in detail for the
process $\nu_e +e \leftrightarrow \nu_e+e$. The incorporation of the 
electron thermal mass alone leads to a 0.1\% increase of $T_d$.
When the thermal phase space in the Born approximation is
also considered, the decoupling temperature further increases.
The total thermal effect
is an increase of $T_d$ by 4.4\%, and the actual
value of the decoupling temperature is
$T_d(\mbox{thermal effects})=1.41$ MeV. 
In conclusion, it is still valid even 
in the presence of a heat bath
that neutrinos are totally decoupled at the time of
$e^\pm$ entropy transfer.

\vspace{1cm}
{\bf Acknowledgements.}
NF gratefully acknowledges a fellowship from the Istituto Nazionale
di Fisica Nucleare, Italy.

\newpage
\begin{center}
\begin{large}
FIGURE CAPTIONS
\end{large}
\vspace{5mm}\
\end{center}
\begin{itemize}
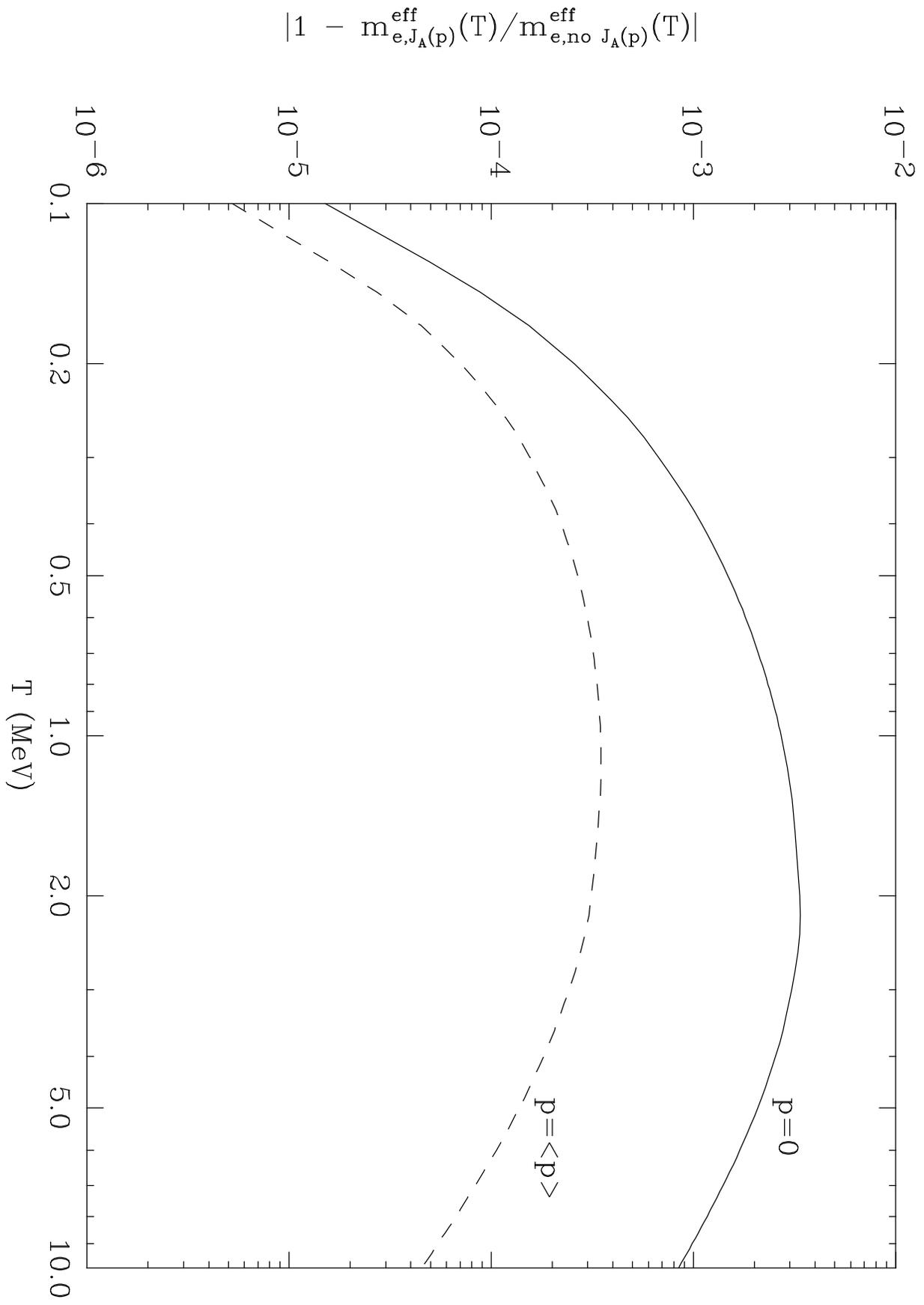
\item [{\rm FIG. 1}] 
The deviation of the effective 
mass of the electron $m_e^{\rm eff}$ 
calculated by including the $J_A(p)$ term from the one without
the $J_A(p)$ term as a function of $T$. The solid line
refers to the electron momentum $p=0$, the dashed line to
the mean value $p = \langle p \rangle$.

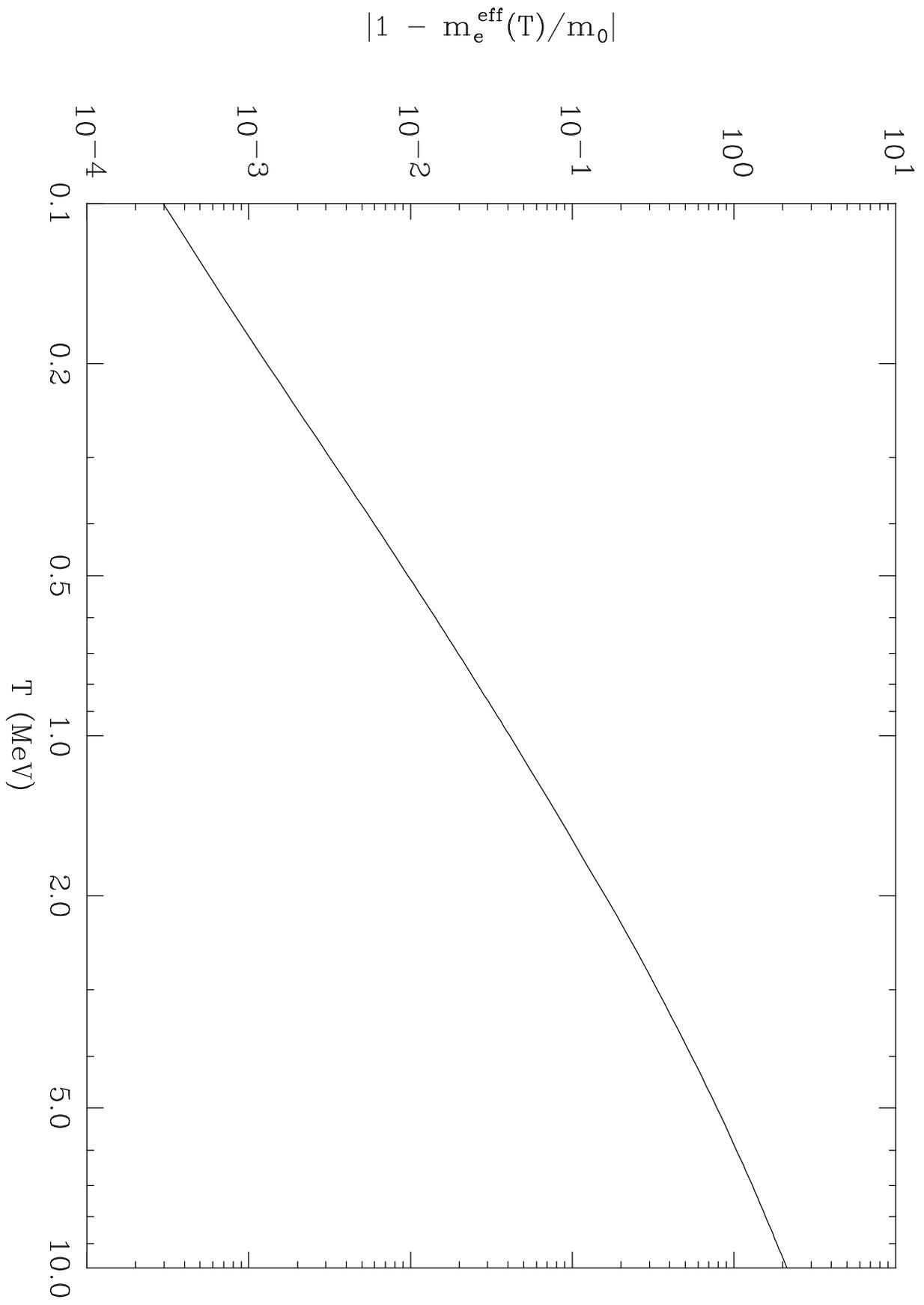
\item [{\rm FIG. 2}] 
The deviation of the thermal electron mass $m^{eff}_e(T)$
from $m_0$ as a function of $T$.

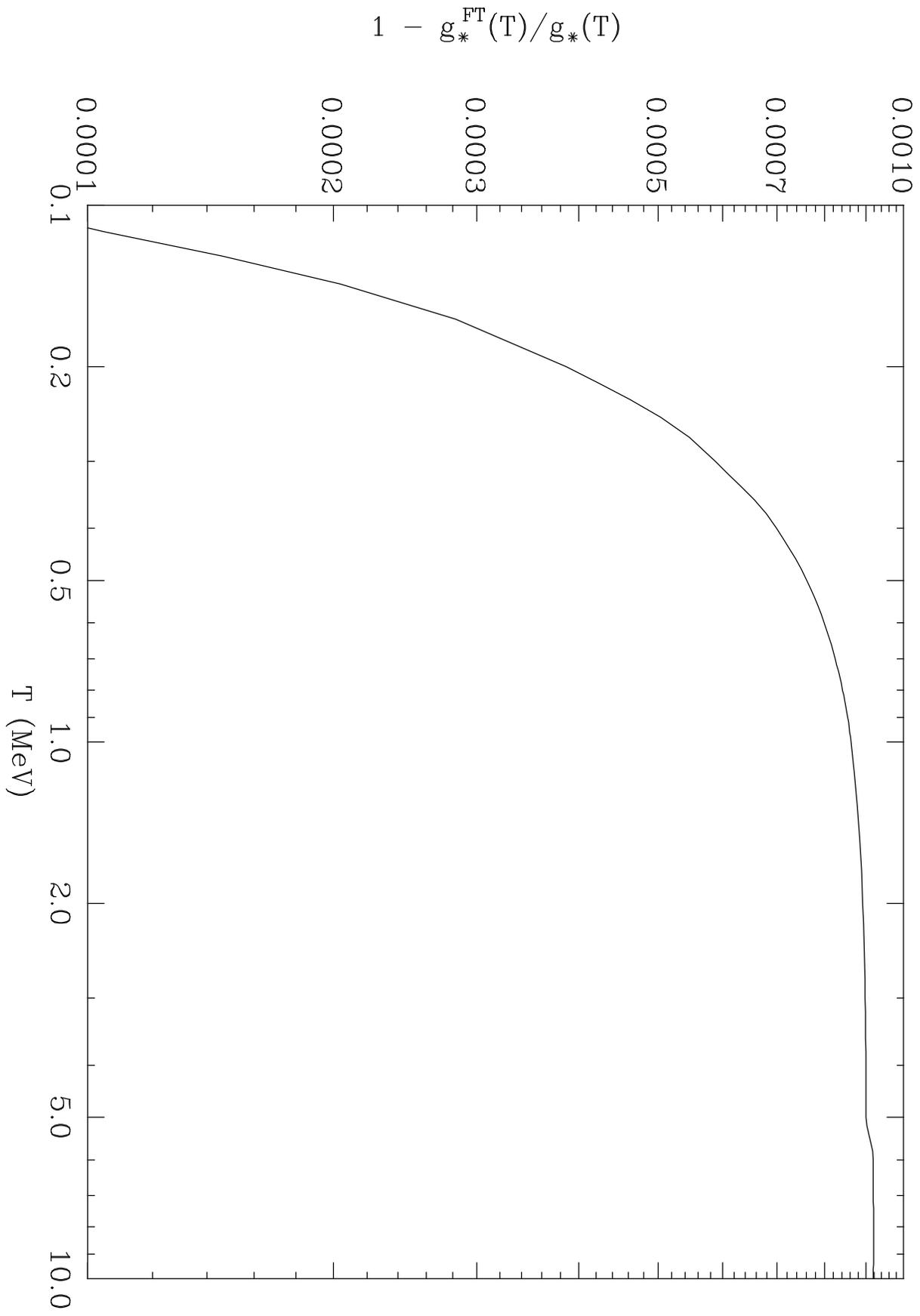
\item [{\rm FIG. 3}] 
The deviation of the number of degrees of freedom with
finite temperature corrections $g_*^{FT}(T)$
from $g_*(T)$ as a function of $T$.

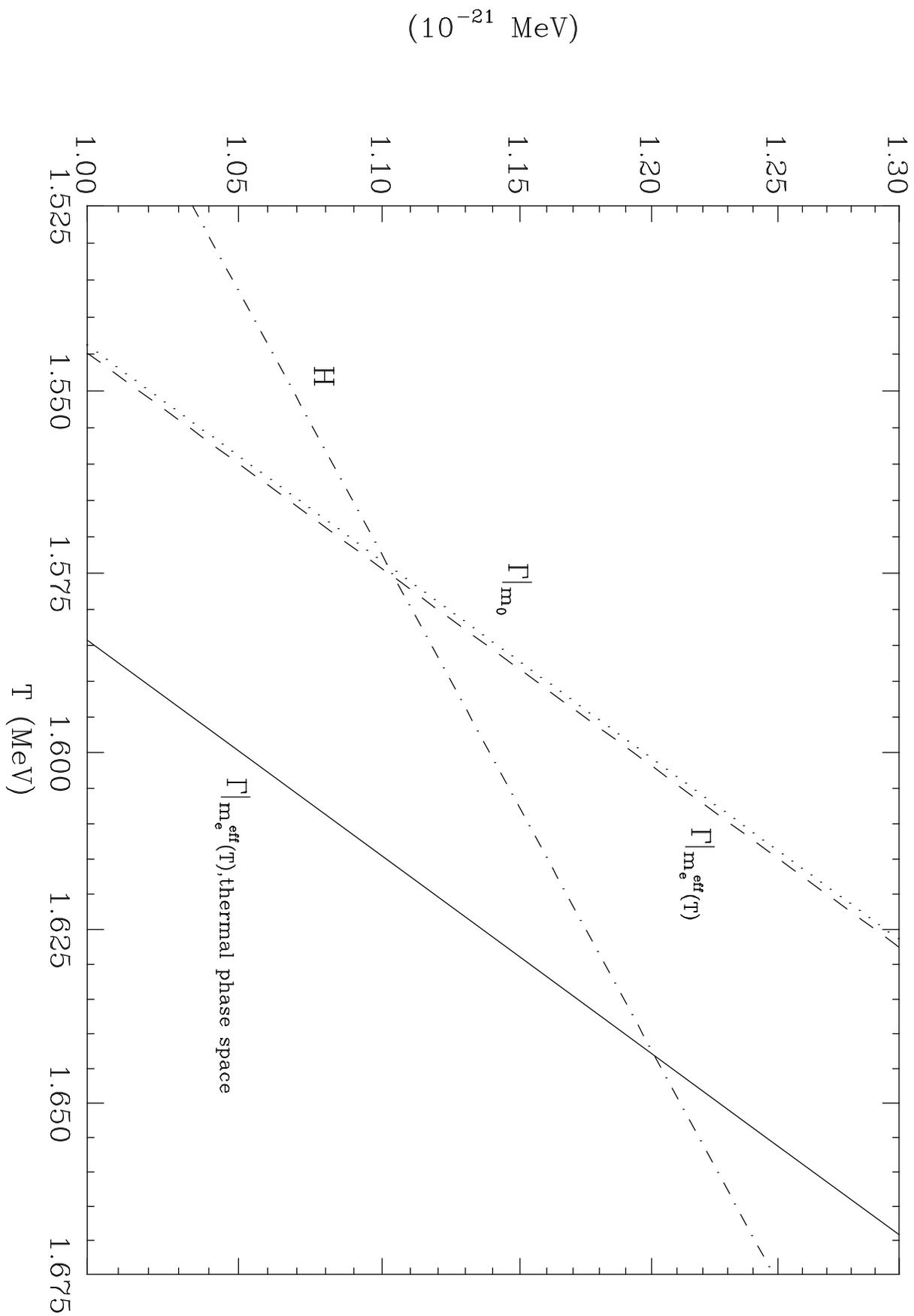
\item [{\rm FIG. 4}] 
The expansion rate $H$ (dash--dotted line) and the interaction rate 
$\Gamma$ for the reaction $\nu_e + e^- \longleftrightarrow \nu_e + e^-$
as functions of $T$, in units of $10^{-21}$ MeV. The
dotted line refers to $\Gamma$ when the thermal
corrections are neglected. The dashed line
represents the case with the electron thermal mass.
The solid line includes both the electron
thermal mass and the thermal phase space.

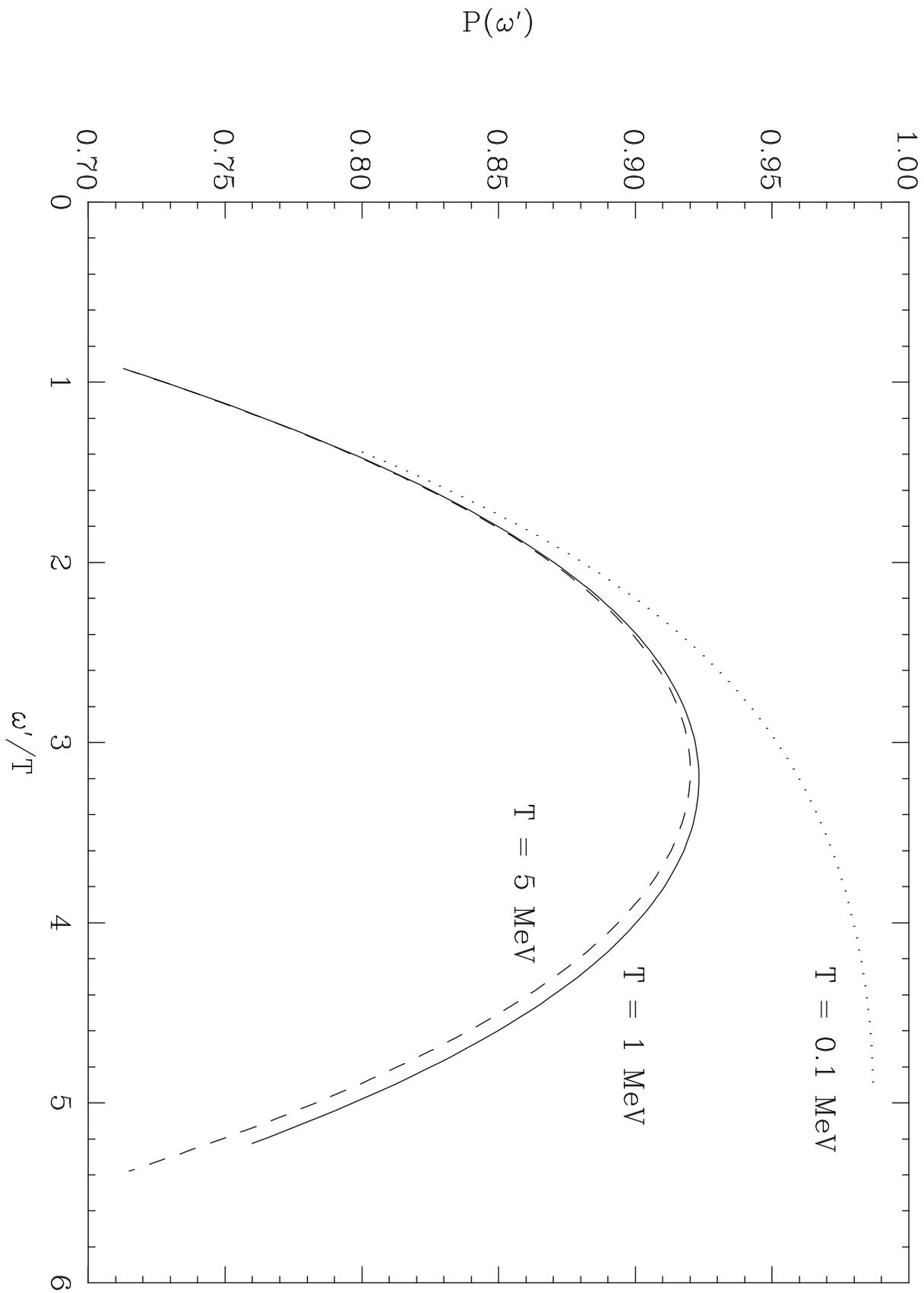
\item [{\rm FIG. 5}] 
The phase space reduction function $P(\omega')$ plotted as a function 
of the final--state neutrino energy $\omega'$ and calculated for the
mean values $E=\langle E \rangle$ and $w=\langle w \rangle$ of the
initial--state energies with $\cos\theta = 0$. The three curves refer to
different values of $T$: $T=0.1$ MeV (dotted line),
$T=1$ MeV (solid line) and $T=5$ MeV (dashed line).

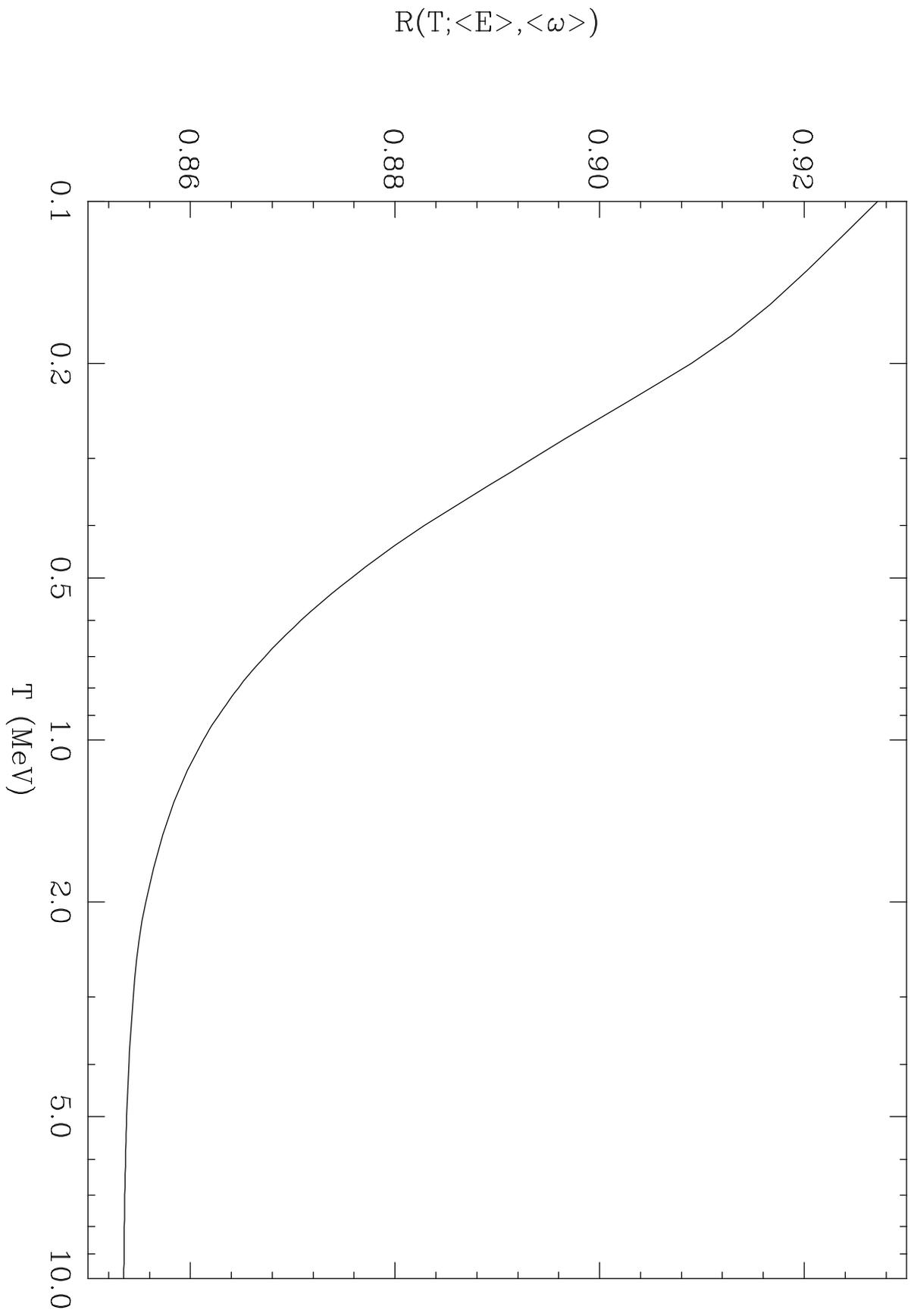
\item [{\rm FIG. 6}] 
The thermal phase space reduction factor $R(T;E,\omega)$ as a function
of $T$, calculated for $E =\langle E \rangle$ and 
$\omega =\langle \omega \rangle$. 

\end{itemize}
\end{document}